\documentclass[aps,superscriptaddress,showpacs,nofootinbib,eqsecnum,prd,notitlepage]{revtex4} 

\usepackage{epsf}  
\usepackage{eucal} 
\usepackage{amssymb}
\usepackage{amsmath} 
\usepackage{graphicx} 
\usepackage{bm}
\usepackage{dcolumn} 
\usepackage{epsfig} 
\usepackage{multirow}
\usepackage{hyperref}
\usepackage{subfigure}

\begin{document}

\title{Casimir force due to condensed vortices in a plane}

\author{J. F. de Medeiros Neto} \email{jfmn@ufpa.br} 
\affiliation{Faculdade de F\'{\i}sica, Universidade Federal do Par\'a, 66075-110, Bel\'em, PA,  Brazil}

\author{Rudnei O. Ramos}\email{rudnei@uerj.br}
\affiliation{Departamento de F\'{\i}sica Te\'orica, Universidade do Estado do Rio de Janeiro, 20550-013, Rio de Janeiro, RJ,  Brazil}

\author{Carlos Rafael M. Santos} \email{carlos.santos@icen.ufpa.br}
\affiliation{Faculdade de F\'{\i}sica, Universidade Federal do Par\'a,
  66075-110, Bel\'em, PA,  Brazil}

\begin{abstract}
  
The Casimir force between parallel lines in a theory describing
condensed  vortices in a plane is determined.  We make use of the
relation between a Chern-Simons-Higgs model and its  dualized version,
which is expressed in terms of a dual gauge field and a vortex
field. The dual model can have a phase of condensed vortices, and, in
this phase, there is a mapping to a model of two noninteracting
massive  scalar fields from which the Casimir force is readily
obtained.  We also discuss the details concerning the boundary
conditions required for  the scalar fields and their association with
those for the vectorial field.  We show that this association is
subtle for the case of the transformations considered.
 
\end{abstract}


\maketitle

\section{Introduction}
\label{intro}

The Casimir effect is a manifestation of the quantum vacuum
fluctuations that can be tested at mesoscopic scales.  This quantum
phenomenon has been of interest to fundamental physics since its
prediction by Casimir in 1948~\cite{casimir}, and it has been studied
extensively both theoretically and experimentally since then (for some
recent reviews on both theory and experiments,  see,
e.g., Refs.~\cite{review,Klimchitskaya:2009cw}).  In particular, many
experiments have been  measuring the Casimir force with increasing
precision. It is then of increasing interest to look for possible
novel situations where the theoretically computed Casimir force can be
confronted to experiments and where related quantum phenomena,
associated with the quantum vacuum, can then be probed and  tested in
the laboratory. 

In this work, we want to study how a vacuum state made of topological
excitations, more precisely, a vacuum constituted of condensed
vortices, will affect the Casimir force between perfectly conducting
parallel lines in a plane.  Let us recall that stable vortex
configurations can appear in important condensed matter systems, like
in high-temperature superconductors and superfluids (for a detailed
presentation, see, e.g., Ref.~\cite{kleinert} and references therein).
There has being an increasing use of superconducting materials to
study the Casimir effect (see, for example, \cite{Bimonte1}).
It has also been pointed out in Ref.~\cite{AlanTDorsey} that unusual
behaviors of superconductors may be found when the sizes of the
samples shrink.  But we note that this is precisely the case, at the
nanometer scales, that we expect that the Casimir force to become more
appreciable.  In the case of vortex-based superconducting
detectors~\cite{AMKadin,AlexeiDSemenovaEtAl_1},
for instance,  it can be expected that the Casimir effect can possibly
alter the microscopic parameters  of the detector, analogously to the
case reported in Ref.~\cite{J.OfLowTemp.Phys.151}.  Since
superconductors can naturally form condensed phases of vortices, it
becomes a matter of interest to investigate how a vacuum state
constituted of a condensate of vortex excitations would affect the
Casimir force. A vortex condensed phase constitutes a particular
example of a nontrivial vacuum state. The Casimir effect being a
manifestation of the quantum vacuum, it is then a fundamental problem
to investigate how a vacuum  state with topological excitations can
affect the Casimir force.

The simplest and, in our opinion, the most direct way for studying a
vortex condensate state is through the use of dual transformations
involving the field variables of the original Lagrangian density.
By following this procedure, we can make explicit the system's
topological excitations content. In the problem that we study in this work,
these topological excitations will be vortex ones.   The duality
transformations are reminiscent of similar  approaches first
used in condensed matter studies performed on the
lattice~\cite{lattice} and of routine use since then.  Through a
series of appropriate dual transformations involving the original
fields in the functional action,  an equivalent action is obtained,
in which the vortex excitations are made explicit.  By properly matching
our dual action to a field theory model, it is then possible to write
it in terms of a vortex field coupled to a vectorial field (for
earlier implementations of this procedure, see for example, the work
done in Refs.~\cite{earlyworks,Kim-Lee,RudneiFelipe}, and
references therein).

Here we investigate the Casimir force for a massive vectorial field in
a Maxwell-Proca-Chern-Simons (MPCS) model. 
{}Following the work in Refs.~\cite{Kim-Lee,RudneiFelipe}, 
we show that this model can be  seen as the dualized
version of a Chern-Simons-Higgs (CSH) model, in which the vortex
excitations of the CSH model are made explicit and considered in a 
vacuum state.  Vortex condensation
in Chern-Simons (CS) type theories, particularly in self-dual models,
have been shown possible for some critical value of the Chern-Simons
parameter~\cite{vortexcond1,vortexcond2}, with the determination of
the condensation  point explicitly obtained in
\cite{RudneiFelipe}. The Casimir force for the dual MPCS type of model
is studied here in this context, deep inside the vortex
condensate phase.

Irrespective of the connection of the dual MPCS model with the CSH
model,  the study done in this work has an interest of its own, which
is associated with the determination of the Casimir force for massive
vectorial fields. Recall that the MPCS model represents, by itself,
massive photons in 2+1 dimensions, with the photon mass having
contributions from  the usual Proca and the CS terms. While the mass
contribution coming from the Proca term can be seen as having been  generated
through a symmetry breaking scalar field term, the contribution from the
Chern-Simons term is of purely topological origin~\cite{reviewCS}.
The issue of the Casimir force for a vectorial field is closely
related to important questions, from both experimental and theoretical
points of views.   {}For instance, in the case of 3+1 dimensions, the
authors of Ref.~\cite{Nature311-336-339}  have analyzed the existence
of new expressions for the electromagnetic field between conducting
plates, where the photon has a possibly non-null mass. They then study
the dependence of the Casimir force with the photon mass. Later, in
Ref.~\cite{AnnalsOfPhys162-1985}, the Proca equations were used to
represent the photon mass. In Ref.~\cite{J.OfLowTemp.Phys.151}, it was
considered the mass  acquired by the photon due to the spontaneous
symmetry breaking that takes place when a superconducting detector
passes from its normal (N) to the superconducting (S) state, as a
consequence of the detection of an external photon. In that reference,
it was argued that  the Casimir effect can alter the S-N transition in
a detectable way and to be able to alter the microscopic parameters of
the detector. Also considered in \cite{J.OfLowTemp.Phys.151} was
the viability of describing the Casimir force  when the corresponding
Maxwell equations are replaced by the Proca ones for massive photons.

The Casimir force between perfectly conducting parallel lines in a
plane for a MCS model has been determined previously
in~\cite{Milton_1990,Milton_1992,Chines1,DanEdneyFelSil}.  In
particular, it has already been shown in Ref.~\cite{Milton_1990} that
the Casimir force obtained in the MCS model is identical to that
derived from a massive noninteracting scalar field in
2+1 dimensions~\cite{JAmbjorn}.  This result can be understood from
the fact that in both theories the field satisfies the Klein-Gordon
equation of motion and both have only 1 degree of freedom. Note that,
in this case,  the Chern-Simons term provides a mass term for the
gauge field, but this is a topological mass that still  maintains the
field with only one (transverse) polarization degree of freedom.
Besides, the boundary conditions (BCs) in both models can be matched,
making the Casimir force in both models to agree.  On the other hand,
in a symmetry-broken case, a Proca mass term is  generated for the
gauge field, which acquires a longitudinal
polarization degree of freedom, in addition to the transverse
polarization.   {}For the case of the MPCS theory, the gauge field now
has two polarization degrees of freedom. Similarly to the case of the
MCS theory, we now expect that the respective Casimir force could be
related to {\it two} massive noninteracting scalar fields.  In fact,
it is well known that in this case the quantum mechanical analogue of
the MPCS theory is equivalent to two noninteracting harmonic
oscillators with distinct frequencies~\cite{reviewCS}. At the quantum
field theory level, this fact must then correspond to the case of two
noninteracting massive scalar fields. This has been shown explicitly
in~\cite{Shiang}, where a mapping between the two theories was
constructed.  However, in order to associate the corresponding Casimir
forces for both  theories, a careful consideration of the boundary
conditions must be accounted for.  This issue was earlier discussed in
Refs. \cite{JAmbjorn,ASGoldhaber}.    Here we will give a detailed
account for the issue of the boundary conditions when mapping our dual
MPCS theory with vortices with a model corresponding to  two
noninteracting massive scalar fields.  This will allow us then to
readily obtain the Casimir force for the model we are studying here.

The remainder of this work is organized as follows. In
Sec. \ref{sec2}, we introduce the MPCS model as a particular limit of
the vortex model  considered in Ref.~\cite{RudneiFelipe} and summarize
the relevant equations and relations that will be of relevance for
this work. In Sec. \ref{sec3},  we discuss the mapping that leads from
the initial MPCS theory to a model of two massive and noninteracting
scalar fields. We also analyze the respective mapping between the
boundary conditions  needed for those two models. In Sec. \ref{sec4},
we then derive the Casimir force related to a vacuum state of
condensed vortex excitations from the dual MPCS theory considered and
contrast the result with the case where vortex excitations are absent.
In Sec. \ref{sec5}, we give our concluding remarks and discuss
possible extensions of our work.  {}Finally, in the Appendix, we give
some technical details.


\section{The MPCS theory as a dual model for vortices in a plane}
\label{sec2}

Let us initially consider the CSH model in 2+1 dimensions, written in
terms of a complex scalar field and an  Abelian gauge field, which
here we will  represent them by $\eta$ and $h_\mu$, respectively. The
quantum partition function and the  action of the model have the forms
(in Euclidean space-time and with  indices running from 1 to 3)

\begin{equation}
Z = \int \mathcal{D}h_{\mu}\mathcal{D}\eta \mathcal{D}\eta^* \exp
\left\{ - S_E[h_\mu, \eta,\eta^*] \right\}\;,
\end{equation}

\begin{equation}
S_E[h_\mu, \eta,\eta^*] = \int d^3 x \left[ - i\frac{\Theta}{4}
  {}\epsilon_{\mu \nu \gamma}h_{\mu}{}H_{\nu \gamma} + |D_\mu \eta|^2
  + V(|\eta|) \right] \;,
\label{actionE}
\end{equation}
with $H_{\mu \nu} = \partial_\mu h_\nu - \partial_\nu h_\mu$, $D_\mu
\equiv \partial_\mu + ieh_\mu$ and $\Theta$ is the CS parameter.
$V(|\eta|)$ is a symmetry breaking polynomial potential, independent
of the phase of the complex scalar field and has a non-null vacuum
expectation value (VEV) $|\langle \eta \rangle| =\nu\neq 0$. As
examples of $V(|\eta|)$, we can cite the usual quartic order potential
$V(|\eta|)=\lambda\left( |\eta|^2 - \nu^2\right)^2/4$ and the
sixth-order self-dual potential~\cite{Jackiw:1990aw}:
$V(|\eta|)=e^4\left( |\eta|^2 -\nu^2\right)^2|\eta|^2/\Theta^2$.
By writing $\eta$ in a polar form $\eta = (\rho/\sqrt{2})\exp{(i\chi)}$,
the VEV for $\eta$ becomes $\nu = \rho_0/\sqrt{2}$.

The field equations associated with $h_\mu$ and $\eta$ are known to
have a nontrivial solution associated with a vortex field
configuration~\cite{Jackiw:1990aw}.  When expressed in polar
coordinates ($r,\chi$), the nontrivial solution can be put in the
generic form that represents charged vortices:

\begin{eqnarray}
\eta_{\rm vortex} &=& \xi(r) \,\exp(i n \chi)\;,
\label{eta vortex}
\\ h_{\mu, {\rm vortex}}  &=& \frac{n}{e} h(r)\; \partial_\mu \chi\;,
\label{h vortex}
\end{eqnarray}
where $n$ is an integer that can be interpreted as the vortex
topological charge and ($\xi(r)$, $h(r)$) are obtained by numerically
by solving the classical field differential equations, subjected to
the BCs:

\begin{eqnarray}
\lim_{r \to 0}\xi(r) = 0, \ \ \ \lim_{r \to \infty}\xi(r) = \nu,
\\ \lim_{r \to 0}h(r) = 0,\ \ \ \lim_{r \to \infty}h(r) = 1. 
\end{eqnarray}
A vortex represented by Eqs. (\ref{eta vortex}) and (\ref{h vortex})
can be seen as carrying an ``electric'' charge $Q$ (the spatial
integral of the 0 component of the density current $j_{\mu}$) attached
to a magnetic flux $\Phi$ given by

\begin{equation}
\Phi \equiv \int d^2x H_{12} = \frac{Q}{\Theta}\;.
\end{equation}
This fact is a direct consequence of the presence of the Chern-Simons
term and also implies in an anyonic behavior of the charge-flux
composite, which has spin $s= Q \Phi/(4 \pi)$~\cite{anyons}. It can
also be demonstrated that when $r$ approaches infinity (or when it is
sufficiently far from the vortex core), the flux $\Phi$ becomes
quantized. $\Phi$ in this case is given by an integer multiple of flux
quantum~\cite{anyons}: $\Phi =2 \pi n/e$.

The vortex degrees of freedom present in the original theory
Eq.~\ref{actionE} can be made explicit through a series of duality
transformations~\cite{Kim-Lee,RudneiFelipe}.  The final result is a
theory of the form of a Maxwell-Chern-Simons-Higgs (MCSH) model, where
the vortex solutions, represented by  Eqs. (\ref{eta vortex}) and
(\ref{h vortex}), are associated with particles represented by a
complex scalar field $\psi$ that is coupled  to a dual vector field
$A_{\mu}$. The original fields and the dual fields, at the  classical
level, are related to each other e.g. by $\rho^2(\partial_\mu \chi + e
h_\mu) = (\sigma/e) \epsilon_{\mu \nu \gamma} \partial_\mu A_\gamma/(2
\pi)$, where $\sigma$ is an arbitrary parameter with mass dimension.
The resulting dual Euclidean action becomes equivalent to a MCSH
theory of  the form~\cite{Kim-Lee,RudneiFelipe}:

\begin{equation}
\label{S_ini}
S_{\rm dual} = \int d^3 x  \left[ \frac{\sigma^2}{16 \pi^2 e^2
    \rho_0^2}F_{\mu\nu}^2 +  i \frac{\sigma^2}{8 \pi^2 \Theta}
  \epsilon_{\mu \nu \gamma} A_\mu \partial_\nu A_\gamma +  \left|
  \partial_{\mu}\psi + i\frac{2\sigma}{e} A_{\mu}\psi \right|^2 +
  V_{\rm vortex}(|\psi|) + {\cal L}_{G} \right] \;,
\end{equation}
where $F_{\mu\nu} = \partial_{\mu}A_{\nu} - \partial_{\nu}A_{\mu}$,
$V(|\psi|)$ is the effective potential term for the vortex field and
${\cal L}_{G}$ is a gauge fixing term. Note that in the dual model,
Eq.~(\ref{S_ini}), the new CS coefficient  appears inversely
proportional to the initial one in Eq.~(\ref{actionE}),  $\Theta \to
-1/(4 \pi^2 \Theta)$.  This dualization of the CS coefficient is a
consequence of the transformations used (see also
Refs.~\cite{Wen:1988uw,Burgess:2000kj}).

As argued in Refs.~\cite{vortexcond1,vortexcond2}, there is a critical
value for the CS coefficient in the CSH theory, below which vortices
are expected to be energetically favorable to condense. In terms of
the dual action~(\ref{S_ini}), this can be expressed in terms of an
existence condition for a VEV for the dual vortex field, given in
terms of the first derivative of the potential with respect to the
vortex field,  $V'_{\rm vortex}(|\psi|=\psi_0/\sqrt{2})=0$, or,
analogously, that the quadratic mass term in the vortex potential be
negative below some critical $\Theta_c$, with $\Theta_c$ determined by
the condition on the second derivative of the effective vortex
potential with  respect to the vortex field, $V''_{\rm
  vortex}(\Theta=\Theta_c)=0$. In Ref.~\cite{RudneiFelipe} this
critical value has been obtained as given by $\Theta_c \simeq
(e^2/\pi) \ln 6$ and shown to be robust against quantum corrections,
changing by no more than about $17\%$.   In this work we are
interested in deriving the Casimir force starting from the dual
action~(\ref{S_ini}) considering the case in which 
vortex condensation is
favorable, {\it i.e}, for the region of parameters where $\Theta <
\Theta_c$.

Since the  Casimir force is related to quantum vacuum fluctuation of
fields, if we want to determine an expression for that force in the
case of the MCSH theory of the form of Eq. (\ref{S_ini}), we can, as
an approximation, consider only small variations of the vortex field
around its nontrivial constant VEV $\psi_0$.  In other words, if we
are deep inside the vortex condensed phase, fluctuations of the vortex
field can be neglected, much like in the London approximation in
condensed matter problems~\cite{kleinert}.  This approximation can
then be seen as a limiting case of Eq.~(\ref{S_ini}), in which the
term $\left| \partial_{\mu}\psi + 2i\sigma A_{\mu}\psi/e \right|^2$ in
Eq. (\ref{S_ini}) gives rise to a Proca-like term ({\it i.e.}, we are
in the vortex symmetry-broken phase in the dual action), which will be
written as $m^{2}A^{\mu}A_{\mu}$.  We can also make use of the
arbitrariness of $\sigma$ to rewrite Eq.~(\ref{S_ini}) in the form of
a MPCS model (see also the Appendix).  Considering $\sigma \equiv 2\pi
e \rho_0$ and going back to Minkowski space-time,  the corresponding
MPCS Lagrangian density then becomes,

\begin{equation}
\label{LMPCS}
\mathcal{L}= - \frac{1}{4} F^{\mu \nu} F_{\mu \nu} +
\frac{1}{2}m^{2}A^{\mu}A_{\mu}+ \frac{\mu}{4}\epsilon^{\mu \nu
  \lambda} A_{\mu}\partial_{\nu}A_{\lambda}.
\end{equation}
where in Eq.~(\ref{LMPCS}), for convenience, we have redefined the
parameters as

\begin{eqnarray}
\mu \equiv  2 e^2 \rho_0^2/\Theta,\;\;\;  m  \equiv  4\pi
\rho_0\psi_0\;.
\label{m}
\end{eqnarray}	

The association of a covariant derivative of a field with a mass term
for a boson, in the broken vacuum state (for the dual theory in our
case),  is a well known result. This is very similar to the mechanism
of mass generation for photons inside superconductors, which can be
explained in terms of a symmetry breaking in the Landau-Ginzburg model
for superconductivity \cite{PWAnderson}. 

In the next section, the Casimir force related to the theory described
by Eq. (\ref{LMPCS}) is determined by noticing that it can be mapped
to an equivalent model of two noninteracting massive scalar fields,
as mentioned in the introduction,  and by choosing the appropriate
boundary conditions for the corresponding scalar fields and the dual
vector field.   In the association between the two theories, the two
initial massive degrees of freedom of the MPCS model are transposed to
two degrees of freedom represented by the scalar fields, as it should
be expected~\cite{reviewCS}.


\section{The Equivalent Model and the Mapping Between the Boundary Conditions}
\label{sec3}

The MPCS theory given by  Eq.~(\ref{LMPCS}) can be mapped, after a
sequence of mathematical transformations, in a model of two
noninteracting massive real scalar fields ($\phi$ and $\varphi$) in
2+1 dimensions~\cite{Shiang}. Next we will explain the main steps
needed for this mapping and that will be useful for setting the
respective BCs needed in the calculation of the Casimir force.

\subsection{The MPCS theory equivalence to two noninteracting scalar fields}

{}From Eq. (\ref{LMPCS}), the Euler-Lagrange equation for the dual
gauge field $A_\mu$ is

\begin{equation}
\label{eqproca}
\partial _{\mu }F^{\mu \alpha }+m^{2}A^{\alpha }+\frac{\mu
}{2}\epsilon^{\alpha \rho \beta }F_{\rho \beta }=0,
\end{equation}
while the canonical momenta are

\begin{equation}
\label{piS}
\pi^{0}=\frac{\partial{\mathcal{L}}}{\partial \dot{A}^{0}} = 0, \;\;\;
\pi^{i}=\frac{\partial{\mathcal{L}}}{\partial
  \dot{A}^{i}}=F^{i0}+\frac{\mu}{2}\epsilon^{ij}A_{j},
\end{equation}
where the indexes $i$ and $j$ vary from 1 to 2. The relation
$\pi^{0}=0$ is a primary constraint of the model, which also shows a
secondary one, given by

\begin{equation}
\label{constraint}
\partial_{i}\pi^{i} + \frac{\mu}{4}\epsilon^{ij}F_{ij} +
m^{2}A_{0}\approx 0.
\end{equation}

The primary and the secondary constraints are directly related to the
reduction of the number of degrees of freedom of the system (from 3 to
2). We also note that the secondary constraint permits one to
write $A_{0}$ in terms of the components $A_i$. This possibility can
be seen as a direct consequence of the fact that the vectorial field
mass $m$ is non-null. As a consequence of the constraints, the
physical degrees of freedom of the system are represented by $A_i$ and
$\pi_i$. The quantum partition function can now be written in the form

\begin{equation}
Z = \int \mathcal{D}A_{i}\ \mathcal{D}\pi^{i}\exp \left[ i\int d^3x
  \left(\pi^{i}\dot{A}_{i} -  \mathcal{H}\right) \right]\;,
\end{equation}
where $\mathcal{H}$ is the physical Hamiltonian density,

\begin{equation}
\label{Hamimiltoniana1}
\mathcal{H}=\frac{1}{2}\pi^{i}K_{ij}\pi^{j}+\pi^{i}Q_{ij}A^{j}+A_{i}S^{ij}A_{j},
\end{equation}
where $K_{ij}$, $Q_{ij}$ and $ S^{ij}$ are defined, respectively, by

\begin{equation}
K_{ij}=g_{ij} +\frac{\partial i\partial j}{m^{2}},
\label{op1} 
\end{equation}

\begin{equation}
Q_{ij}=\frac{\mu}{2}\left(\epsilon_{ij}+
\frac{1}{m^{2}}\partial_{i}\tilde{\partial_{j}}\right),
\ \ \ \ \ \tilde{\partial_{i}}=\epsilon_{ij}\partial^{j}, 
\label{op1B} 
\end{equation}

\begin{equation}
S^{ij}=\frac{1}{2}\left(1+\frac{\mu^{2}}{4m^{2}}\right)
\left[\left(\nabla^{2}-m^{2}\right)g^{ij} +
  \partial^{i}\partial^{j}\right].
\label{op2}
\end{equation}

It is important to note that in order to write the Hamiltonian density
in the form  Eq.~(\ref{Hamimiltoniana1}),  the surface terms generated
by the integrals of $\partial_i(\pi^i\partial_j\pi^j)$,
$\partial_i(\pi^i \epsilon^{jk} F_{jk})$, $\partial^i(A^j\partial_i
A_j)$ and $\partial^i(A^j\partial_j A_i)$ are neglected. As we will
show below, this can be shown to be indeed the case for the boundary
conditions considered here.

Next, we introduce two new variables, $\tilde{A}_i$ and
$\tilde{\pi}^{i}$ ($i$ = 1, 2), defined by the relations~\cite{Shiang}

\begin{eqnarray}
A_{1} & = & \left({\hat{\cal{O}}_1}^{-1}\tilde{A}_{1} -
{\hat{\cal{O}}_2}^{-1}\tilde{A}_{2}\right)/(2\theta),  
\label{Trans1}\\ 
A_{2} & = &
\hat{\cal{O}}_1\tilde{\pi}^{1}+\hat{\cal{O}}_2\tilde{\pi}^{2}, 
\label{Trans2}\\  
\pi^{1} & = & \theta\hat{\cal{O}}_1\tilde{\pi}^{1} - \theta
\hat{\cal{O}}_2\tilde{\pi}^{2}, \label{Trans3}\\  \pi^{2} & = &
-\left({\hat{\cal{O}}_1}^{-1}\tilde{A_{1}} +
{\hat{\cal{O}}_2}^{-1}\tilde{A_{2}}\right)/2. \label{Trans4}
\end{eqnarray}
where 

\begin{equation}
\theta=\sqrt{ m^{2}+\frac{\mu^{2}}{4}}\;,
\label{theta}
\end{equation}
and $\hat{\cal{O}}_1$ and $\hat{\cal{O}}_2$ are operators whose
squares are given, respectively, by

\begin{equation}
\hat{\cal{O}}_1^{2}=\left( -\frac{1}{2}\theta^{2}K_{11}-\theta Q_{12}+
S^{22} \right)^{-1},
\label{Operator1}
\end{equation}

\begin{equation}
\hat{\cal{O}}_2^{2}=\left( -\frac{1}{2}\theta^{2}K_{11}+\theta Q_{12}+
S^{22} \right)^{-1}.
\label{Operator2}
\end{equation} 
We note from the above equations that when acting the operators
$\hat{\cal{O}}_1$ and $\hat{\cal{O}}_2$ on some function
(e.g. $\phi(x)$), they cannot be simply written in terms of the
derivatives of the function. In Eqs.~(\ref{Trans1}) and
(\ref{Trans4}), $\tilde{\pi}^{i}$ and $\tilde{A_{i}}$ can be seen as
intermediate variables, related to the fields $\left\{\phi,
\varphi\right\}$ and their respective momenta $\left\{\pi_{\phi},
\pi_{\varphi}\right\}$, as

\begin{eqnarray}
\tilde{\pi_{1}} & = &
\frac{1}{\sqrt{2}}\pi_{\phi}-\sqrt{2}\left(\frac{S^{12}}{\theta}+
\frac{Q_{22}}{2}\right)\phi,
\label{pi1til}\\
\tilde{\pi_{2}} & = &
\frac{1}{\sqrt{2}}\pi_{\varphi}-\sqrt{2}\left(\frac{-S^{12}}{\theta}+
\frac{Q_{22}}{2}\right)\varphi,
\label{pi2til}\\
\tilde{A_{1}} & = & \sqrt{2}\phi,
\label{A1til}\\
\tilde{A_{2}} & = & \sqrt{2}\varphi.
\label{A2til}
\end{eqnarray}

The set of mathematical transformations shown above makes it possible
to  rewrite the Hamiltonian of the MPCS model as a sum of two
separated and independent Hamiltonians associated with two
noninteracting  scalar fields $\varphi$ and $\phi$, {\it i.e.},

\begin{equation}
\mathcal{H} = \frac{1}{2}\left[\pi_{\phi}^{2} +
  \phi(m_{1}^{2}-\nabla^{2})\phi \right]+
\frac{1}{2}\left[\pi_{\varphi}^{2} +
  \varphi(m_{2}^{2}-\nabla^{2})\varphi \right], 
\label{Hfinal}
\end{equation} 
where 

\begin{equation}
m_{1}=\theta - \frac{\mu}{2}, \ \ \ \  m_{2} = \theta + \frac{\mu}{2}.
\label{m1m2}
\end{equation} 

The relation between the model described by $\mathcal{H}$,
Eq. (\ref{Hfinal}) and the MPCS theory, can now be  used to obtain the
Casimir force for the dual model Eq. (\ref{LMPCS}), describing a
condensed vortex in the dual formalism. Since the Casimir force for a
massive scalar field in 2+1 dimension is well known~\cite{JAmbjorn},
provided well defined BCs are considered, we now turn our attention  to
this issue of setting the BCs for the mapped theory.

\subsection{The BC mapping between the gauge field and the scalar fields}

The method that we use here for determining the Casimir force for the
MPCS model is to associate it with a model of scalar fields, as
explained in the previous subsection. The involved mathematical form
of the mapping between the  vectorial field and the two scalar fields,
however, makes the problem of  fixing the BCs in this case a nontrivial 
one. Below, we will elaborate on this problem of mapping the
required BCs.   As we will show next, some usual BCs considered for
scalar fields in  Casimir problems cannot be directly written in terms
of the vectorial field $A_{\mu}$ (at least in a simple form). This is
an important issue,  since it is well known that the Casimir force
(for both its modulus and  orientation) depends significantly on the
BCs considered. 

Our aim is to obtain the Casimir force for the vectorial field by
equating  it to a sum of two previously known expressions of Casimir
forces for two scalar  fields that have well-posed BCs. To be able to
make this association  between the two models and to use the
corresponding Casimir force result known for massive scalar fields,
the  BCs for the scalar fields have to be related to  well-posed and
physically acceptable BCs for the vectorial field.  As an
illustration, we could wonder whether the condition 
for the fields
$\phi$ and $\varphi$ to vanish at the boundaries,  
which is a well studied BC for scalar fields in Casimir problems,
would or not imply in perfect conductor BCs (for instance) for the
vectorial field  and vice versa. To answer this question requires
having a clear map from  $\left\{\phi, \varphi \right\}$ (and/or the
derivatives of those fields) into $\left\{ A_{0}, A_{1},
A_{2}\right\}$ (and/or the derivatives of those fields components), at
least at the boundaries. Hence, we need to invert the relations
$\left\{ A_{0}, A_{1}, A_{2}\right\} \to \left\{\phi, \varphi
\right\}$ given in in the previous subsection. With this aim, we first
use the expressions for $\tilde{A_{1}}$ and $\tilde{A_{2}}$, given by
Eqs. (\ref{A1til}) and  (\ref{A2til}), and substitute them in
Eq. (\ref{Trans1}). {}From this, we obtain,

\begin{equation} 
\label{A1}
A_{1} =  \left[{\hat{\cal{O}}_1}^{-1}\phi -
  {\hat{\cal{O}}_2}^{-1}\varphi\right]/(\sqrt{2}\theta). 
\end{equation} 
Since the physical BCs are specified in configuration space,
we need to further elaborate on the meaning of the terms
${\hat{\cal{O}}_1}^{-1}\phi$
and ${\hat{\cal{O}}_2}^{-1}\varphi$ appearing in Eq. (\ref{A1}),
in particular at the boundaries.

Let us consider initially the first term in Eq. (\ref{A1}),
${\hat{\cal{O}}_1}^{-1}\phi$.
Using the explicit forms of the
operators $K_{11}$, $Q_{12}$ and $S^{22}$, given in
Eqs. (\ref{op1})-(\ref{op2}), we can  write that

\begin{equation} 
\hat{\cal{O}}_1^{2}\phi = \left(A - B\partial_{1}^{2}\right)^{-1}\phi,
\label{mu1-2}
\end{equation}
where two new constants, $A$ and $B$, have been introduced in the above
equation and they are given, respectively, by

\begin{equation} 
A \equiv \frac{\theta^2}{2} - \frac{\theta\mu}{2} +
\frac{m^2}{2}\left( 1 + \frac{\mu^2}{4m^2} \right) , \ \ \ \ \ \ B
\equiv -\frac{\theta^2}{2m^2} - \frac{\theta\mu}{2m^2} - \frac{1}{2} -
\frac{\mu^2}{4m^2}\;.
\end{equation}
{}From Eq.~(\ref{mu1-2}) we see that ${\hat{\cal{O}}_1}^{-1}\phi$  can
be written as $\left(A - B\partial_{1}^{2}\right)^{1/2}\phi$.  
Let us now evaluate this expression at the boundaries.
Our physical system is constrained in an  infinite
strip, with boundaries at $x = 0$ and $x = a$. 
By also considering
that the field $\phi$ obeys the Neumann BC, with $\partial_{1}\phi(x =
0) = \partial_{1}\phi(x = a) = 0$, thus, {\it at the boundaries}, we
can write

\begin{equation} 
\label{temp1}
\left(A -
B\partial_{1}^{2}\right)\phi=
\left(\sqrt{A} + i\sqrt{B}\partial_{1}\right)^{2}\phi \;.
\end{equation} 
{}From Eqs.~(\ref{mu1-2}) and (\ref{temp1}), we can now write

\begin{equation} 
\hat{\cal{O}}_1^{-2}\phi = \left(A - B\partial_{1}^{2}\right)\phi  =
\left(\sqrt{A} + i\sqrt{B}\partial_{1}\right)^{2}\phi.
\end{equation}
Hence, at the boundaries $x = 0$ and $x = a$, we determine that

\begin{equation} 
\label{O1MenosUmPhi}
\hat{\cal{O}}_1^{-1}\phi = \left(\sqrt{A} +
i\sqrt{B}\partial_{1}\right)\phi = \sqrt{A}\phi.
\end{equation}
With these results, it is now easy to write the first term of the
left-hand side of Eq.~(\ref{A1}) in the configuration (coordinate) space and 
at the boundaries. We
note that, in  Eq.~(\ref{O1MenosUmPhi}), $\sqrt{A}$ does not represent
an eigenvalue  of $\hat{\cal{O}}_1^{-1}$, but the mathematical
expression of that operator  itself (at the boundaries). 

We can use analogous considerations also for
$\hat{\cal{O}}_2^{-1}\varphi$, the second term in the left-hand side of
Eq. (\ref{A1}).  {}From similar arguments as those used for
$\hat{\cal{O}}_1^{-1}\phi$ and  considering Neumann BC for $\varphi$,
we can write, at the boundaries, that 

\begin{equation} 
\label{O2MenosUmVarphi}
\hat{\cal{O}}_2^{-1}\varphi \equiv \sqrt{C}\varphi,
\end{equation}
where the constant $C$ in the above equation is defined as

\begin{equation} 
C \equiv \frac{\theta^2}{2} + \frac{\theta\mu}{2} +
\frac{m^2}{2}\left( 1 + \frac{\mu^2}{4m^2} \right).
\end{equation}

{}From the above results, we can write Eq.~(\ref{A1}), at the
boundaries, as $A_{1} =  \left[\sqrt{A}\phi -
  \sqrt{C}\varphi\right]/(\sqrt{2}\theta)$. Hence we see that $A_{1}$
must also obey the Neumann BC:

\begin{equation}
\partial_{1}A_{1}(x = 0) = \partial_{1}A_{1}(x = a) = 0\;.
\end{equation}

Likewise, we can proceed analogously to obtain the required conditions
for $A_2$. By making use of Eqs. (\ref{Trans2}), (\ref{pi1til}) and
(\ref{pi2til}), we obtain that

\begin{equation}
\label{A2Ini}
A_{2}=\frac{\left( \hat{\cal{O}}_1 \pi_{\phi} +
  \hat{\cal{O}}_2\pi_{\varphi} \right) }{\sqrt{2}} -
\sqrt{2}\hat{\cal{O}}_1\left( \frac{S^{12}}{\theta} +
\frac{Q_{22}}{2}\right)\phi - \sqrt{2}\hat{\cal{O}}_2 \left( -
\frac{S^{12}}{\theta} +  \frac{Q_{22}}{2}\right)\varphi.
\end{equation}
We can now use the Eq.~(\ref{A2Ini}) to determine the behavior of
$A_{2}$ at the boundaries. Using Eqs. (\ref{O1MenosUmPhi}) and
(\ref{O2MenosUmVarphi}), we can write (for $x = 0$ and $x = a$) that
$\hat{\cal{O}}_1\phi = \phi/\sqrt{A}$ and $\hat{\cal{O}}_2\varphi =
\varphi/\sqrt{C}$. Noticing that we are considering the Neumann BC for
$\phi$ and $\varphi$, we can use the Hamilton equations ($\pi_{\phi} =
\partial_0\phi$ and $\pi_{\varphi} = \partial_0\varphi$) and  the
explicit forms of $Q_{22}$ and $S^{12}$ to rewrite Eq.~(\ref{A2Ini})
as

\begin{equation}
\label{A2v2}
A_{2}= \frac{\partial_0\phi}{\sqrt{2A}} +
\frac{\partial_0\varphi}{\sqrt{2C}}.
\end{equation}
Equation~(\ref{A2v2}) implies that $A_{2}$ must also obey the Neumann BC
(since $\phi$ and $\varphi$ are subjected to the same type of BC).

The BCs considered for $A_{1}$ and $A_{2}$, together with the
Euler-Lagrange equations and the definitions of the canonical momenta,
define the components of the strength tensor at the boundaries. The
behavior of those components should not be confused with a new BC
imposed to the vectorial field, but just direct implications of the
Neumann BCs considered for $A_1$ and $A_{2}$. {}For instance, from the
definition of $\pi^{i}$ given in Eq.~(\ref{piS}), we get

\begin{equation}
F^{20} = \pi^2 + \mu A_{1}/2\;,
\end{equation}
or yet, from Eqs. (\ref{Trans1}), (\ref{Trans4}), (\ref{A1til}), and
(\ref{A2til}),

\begin{equation}
F^{20} = \frac{\mu}{2} A_{1}
-\frac{\sqrt{2}}{2}\left[{\hat{\cal{O}}_1}^{-1}\phi +
  {\hat{\cal{O}}_2}^{-1}\varphi\right].
\end{equation}
Thus, at the boundaries and  using Eqs. (\ref{O1MenosUmPhi}) and
(\ref{O2MenosUmVarphi}), we obtain that

\begin{equation}
\label{F20AtBoundaries}
F^{20} = \frac{\mu}{2} A_{1} -\frac{\sqrt{2}}{2}\left[\sqrt{A}\phi +
  \sqrt{C}\varphi\right].
\end{equation}
Since $A_{1}$, $\phi$ and  $\varphi$ are subjected to the Neumann BC,
Eq.~(\ref{F20AtBoundaries}) implies that $F^{20}$ is also subjected to
the same BC. We can also write those BCs in terms of the dual tensor
$F^{\mu}$, defined by $F^{\mu} \equiv \epsilon^{\mu \nu \rho}
\partial_{\nu} A_{\rho}$, to obtain

\begin{equation}
\label{BCforF1}
\partial_1 F^{1}(x = 0) = \partial_1 F^{1}(x = a) = 0.
\end{equation}
The result given by Eq.~(\ref{BCforF1}) can be seen as a BC for the
vectorial field and a direct consequence of the Neumann BCs considered
for $A_1$ and $A_2$, which,  in turn, are a direct consequence of the
Neumann BCs considered for $\phi$ and $\varphi$. We can say that
Eq.~(\ref{BCforF1}) is the analogue of the BC $F^{1}(x = 0) = F^{1}(x
= a) = 0$ considered in Ref.~\cite{Milton_1990}. Also, we note that,
in a similar manner to what occurred in Ref.~\cite{Milton_1990}, the
BC given by Eq.~(\ref{BCforF1}) can be seen as a consequence of the
Bianchi identity $\partial_{\nu} F^{\nu} = 0$, together with the
statics requirement $\partial_0 F^0 = 0$,  imposed to a perfect
conductor. To better see this in a clearer manner,  we can first
evaluate Eq.~(\ref{eqproca}) for $\alpha = 0$ and $\alpha = 1$.  Using
the BCs considered above, we can write (at the boundaries)

\begin{eqnarray}
\label{EulerLagrange_alpha0}
\left( \partial_1\partial^1 + m^2 \right)A_0 - \mu\partial_2 A_1  +
\partial_2 F^{20} & = & 0,\\
\label{EulerLagrange_alpha1}
\left( \partial_0\partial^0 + \partial_2\partial^2 + m^2 \right)A^1 -
\partial_0\partial^1 A_0  + \mu F_{20} & = & 0.
\end{eqnarray}
{}From Eq. (\ref{EulerLagrange_alpha1}), it is easy to see (by taking
the derivative with respect to $x$ and using the BCs) that
$\partial_0\partial_1\partial^1 A_0 = 0$.  Using this result and
representing $A_0$ in terms of its transverse {}Fourier transform  (in
$y$ and $t$) \cite{Milton_1990},

\begin{equation}
A_0(x, y, t) = \int\frac{d\omega}{2\pi}e^{-i\omega t}
\int\frac{dk}{2\pi}e^{ik y}\tilde{A}_0(x, k, \omega),
\end{equation}
we see that the condition $\partial_0\partial_1\partial^1 A_0 = 0$
(valid for any $t$ and $y$), implies that
$\partial_1\partial^1\tilde{A}_0(x, k, \omega) = 0$ and, therefore,
$\partial_1\partial^1 A_0 = 0$ (for $x = 0$ and $x = a$). We can now
use this result in Eq.~(\ref{EulerLagrange_alpha0}) to obtain

\begin{equation}
\label{EulerLagrange_alpha0_v2}
m^2 A_0 - \mu\partial_2 A_1  + \partial_2 F^{20} = 0.
\end{equation}
Since $A_1$ and $F^{20}$ are subjected to the Neumann BC,
Eq.~(\ref{EulerLagrange_alpha0_v2}) implies (by deriving with respect
to $x$) that $A_0$ is also subjected to the same kind of BC as well:
$\partial_1 A_0 = 0$ at $x = 0$ and $x = a$. Hence, at the boundaries,
we have

\begin{equation}
\label{F2-Boundaries}
F^2 = F_{01} = \partial_0 A_1,
\end{equation}
where we made use of the BC for $A_0$. 

By considering that $F^{0}$ is subjected to the statics requirement
$\partial_0 F^0 = 0$ \cite{Milton_1990}, we get (at the boundaries,
where $\partial_1 A_2  = 0$),

\begin{equation}
\label{staticsRequirement}
\partial_0 F^0 = \partial_0 F_{12} = \partial_0\partial_2 A_1  = 0.
\end{equation}

We can now use Eq.~(\ref{staticsRequirement}) to establish the value
of $\partial_2 F^2$ at the boundaries and show that it must vanish as
well.  This result will then be used below, together with the Bianchi
identity, to obtain equally that $\partial_1 F^1 = 0$, which can be
seen as a direct consequence of the Bianchi identity and the statics
requirement.  {}First, we note that since $\partial_1 A_0 = 0$ at the
boundaries, we can write,  for $x = 0$ or $x = a$ that

\begin{equation}
\label{partial2F2}
\partial_2 F^2 = \partial_2 F_{01} = \partial_2\partial_0 A_1.
\end{equation}
By comparing Eqs.~(\ref{partial2F2}) and (\ref{staticsRequirement}),
we see that the statics requirement implies that $\partial_2 F^2=0$ at
the boundaries. Using this condition together with the statics
requirement,  we get likewise that $\partial_1 F^1 = 0$ at the
boundaries. Thus,  the BC $\partial_1 F^1 = 0$ can be seen as a
consequence of the statics  requirement and the Bianchi identity
considered here and in Ref.~\cite{Milton_1990}.

By using the definitions of the canonical momenta and the considerations
about the behavior of $A_i$ at the boundaries, it is easy to prove
that the  surface terms generated by the integrals of
$\partial_i(\pi^i\partial_j\pi^j)$, $\partial_i(\pi^i \epsilon^{jk}
F_{jk})$, $\partial^i(A_j\partial_i A_j)$ and
$\partial^i(A^j\partial_j A_i)$, that appear in the generating
functional, will give no  contributions. This justifies neglecting
those contributions to the partition function, as we have assumed.
Analogously, the BC considered here, written in terms of $\phi$,
$\varphi$ and their respective conjugate momenta allow us to neglect
the surface terms related to those fields in the process of obtaining
the final Hamiltonian density Eq.~(\ref{Hfinal}).


\section{The Casimir force}
\label{sec4}
  
By having the relevant BCs fixed, it becomes straightforward to find
the Casimir force for the dual MPCS theory Eq.~(\ref{LMPCS}).  This
follows directly from  the equivalence between the original theory
Eq.~(\ref{LMPCS}) with the model represented by Eq.~(\ref{Hfinal}).
The Casimir force for a massive scalar field subjected to the Neumann (or
Dirichlet) BC in 2+1 dimensions (which is also the same as the one
computed for a MCS theory)  is \cite{JAmbjorn,Milton_1990}

\begin{equation} 
f_{\rm scalar}(m_s,a)= - \frac{1}{16\pi a^{3}}\int_{2m_s
  a}^{\infty}dy\  \frac{y^{2}}{ e^y - 1} \;,
\label{fs}
\end{equation} 
where $m_s$ is the mass of the scalar field.  The integral in
Eq.~(\ref{fs}) is a second Debye function~\cite{abra},

\begin{equation}
\int_{x}^{\infty}dy\  \frac{y^{2}}{ e^y - 1} = \sum_{k=1}^{\infty}
e^{-k x} \left(\frac{x^2}{k} + 2 \frac{x}{k^2} +  2 \frac{1}{k^3}
\right)\;,
\label{debye}
\end{equation}
indicating that the Casimir force due to massive scalars exponentially
decays with $m_s a$.
 
Using the equivalence between  Eq.~(\ref{LMPCS}) and
Eq.~(\ref{Hfinal}), we can then immediately write the corresponding
Casimir force, in the presence of a vortex condensate,  as

\begin{equation} 
f_{\rm vortex} = f_{\rm scalar}(m_1,a)+ f_{\rm scalar}(m_2,a)\;,
\label{fvortex}
\end{equation} 
where $m_1$ and $m_2$, using Eqs.~(\ref{m}), (\ref{theta}) and
(\ref{m1m2}),  are given by 

\begin{equation}
m_{1(2)} = \frac{e^2 \rho_0^2}{|\Theta|} \left( \sqrt{1 + \frac{16
    \pi^2 \psi_0^2 \Theta^2}{e^4 \rho_0^2} } \mp 1 \right)\;.
\label{m12}
\end{equation}

{}For small values of mass, $m a \lesssim 1$, Eq.~(\ref{fs}) can be
expressed as

\begin{equation}
f_{\rm scalar}(m_s,a)= - \frac{1}{8\pi a^{3}}\left[ \zeta(3) - (a
  m_s)^2 + \frac{2 (a m_s)^3}{3} - \frac{(a m_s)^4}{6} + {\cal O} (a^5
  m_s^5) \right]\;,
\label{fseries}
\end{equation}
where $\zeta(x)$ is the Riemann zeta function.  Using Eq. (\ref{m12})
and keeping for simplicity up to the quadratic term  in the mass in
Eq. (\ref{fseries}), we obtain for the Casimir force
Eq. (\ref{fvortex}) the result

\begin{equation}
f_{\rm vortex}  \simeq - \frac{1}{4\pi a^{3}}\left[ \zeta(3) -
  \left(\frac{e^2 \rho_0^2}{\Theta}\right)^2 a^2 \left( 1 + \frac{8
    \pi^2 \psi_0^2 \Theta^2}{e^4 \rho_0^2} \right)\right]\;.
\label{fvortex2}
\end{equation} 
The result (\ref{fvortex}) allows us to immediately conclude that in
the presence of vortex matter ($\psi_0 \neq 0$), the Casimir force is
always {\it smaller} in magnitude than in the absence of vortices.

There are two mass scales in our original model  Eq.~(\ref{actionE}),
which are the mass for the gauge field $h_\mu$ in the broken  phase,
$m_h$, and the mass for the scalar field $\eta$, $m_\eta$. These
masses  can be related to the relevant scales in the context of
superconductivity. The two naturally occurring length scales in the
theory  of superconductivity are the penetration depth, $\lambda =
1/m_h$, which describes the typical length into which a magnetic field
can penetrate into a superconductor and the coherence length, $\xi =
1/m_\eta$, which describes the length scale at which the order
parameter varies in space. The ratio between these two lengths is the
Ginzburg-Landau parameter, $\kappa = \lambda/\xi \equiv
m_\eta/m_h$. Values of $\kappa > 1/\sqrt{2}$ characterize type-II
superconductors. Type-II superconductors in the presence of a magnetic
field can form a stable vortex state (the Shubnikov
phase~\cite{tink}).  On the other hand, materials with $\kappa <
1/\sqrt{2}$ characterize type-I superconductors.  In type-I
superconductors a magnetic field will destroy superconductivity
without allowing the formation of a stable vortex state.

\begin{figure}[htb]
  \vspace{0.75cm} \epsfig{figure=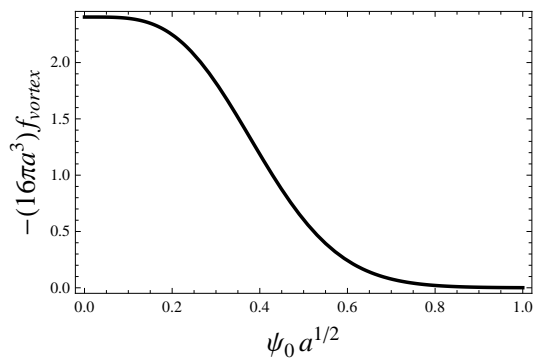,angle=0,width=7cm}
\caption[]{\label{fig1} The Casimir force as a function of the vortex
  condensate $\psi_0$, for the choice of parameters:  $\Theta/e^2=0.1$
  and $\rho_0 a^{1/2} =1$.}
\end{figure}

Using the parameters of the original CSH model Eq.~(\ref{actionE}) and
taking as an example the self-dual potential for the scalar
field~\cite{Jackiw:1990aw}, we have that $m_h = e \rho_0$ and $m_\eta
= e^2 \rho_0^2/\Theta$.  The Ginzburg-Landau parameter becomes $\kappa
= e \rho_0/\Theta$.  As shown in \cite{RudneiFelipe}, vortices are
energetically favored to condense for values of the CS parameter below
a critical value $\Theta_c \approx (e^2/\pi)\ln 6 \simeq 0.57 e^2$ and
for $\Theta < \Theta_c$ we have for the vortex condensate $\psi_0^2
\approx  m_\eta \sqrt{6-\exp(\pi \Theta/e^2)}$. By expressing
Eq.~(\ref{fvortex2}) in terms of these values, we can write the
fractional difference for the Casimir force without vortices, $f_{\rm
  vortex}(\psi_0=0)$, and in the presence of vortices ($\psi_0 \neq
0$) as

\begin{equation}
\frac{\Delta f}{f} \equiv \frac{f_{\rm vortex}(\psi_0=0) - f_{\rm
    vortex}(\psi_0)}{f_{\rm vortex}(\psi_0=0)} \approx \frac{\left(
  m_\eta a \right)^2  8 \pi^2 \frac{\Theta}{e^2} \sqrt{6-\exp(\pi
    \Theta/e^2)} } {\zeta(3) - \left(m_\eta a\right)^2}\;.
\label{fraction}
\end{equation}
If we use representative values consistent with the above requirements
of vortex condensation and in the regime of validity of
Eq.~(\ref{fvortex2}), e.g., $\Theta/e^2 =0.1$ and $m_\eta a=0.1$, we
obtain for the ratio Eq.~(\ref{fraction}) the result $\Delta f/f \simeq
0.14$, representing already a Casimir force that is $14\%$ smaller
due the presence of a vortex condensate.  {}For larger values of
$m_\eta a$, or equivalently for $m_{1(2)} a \gtrsim 1$, we need to
solve numerically for the integral in Eq.~(\ref{fs}), with the
corresponding Casimir force  decreasing exponentially due to the
characteristic second Debye function displayed by the Casimir force
for a massive scalar particle Eq. (\ref{fs}).  In {}Fig. \ref{fig1} we
show the Casimir force Eq. (\ref{fvortex}) as a function of arbitrary
values for the vortex condensate.

The overall decrease of the Casimir force when in the presence of a
vacuum state with  vortices can be interpreted as follows. Vortices
are expected to repel each other, much like as in the standard
mean-field phenomenology for type-II superconductors when vortices can
form~\cite{tink}, e.g.  in the Shubnikov phase, where above some
critical magnetic field vortices are present.  The repelling vortices
will exert an opposite, repulsive force on the external conducting
lines that tend to counterbalance the attractive Casimir force,
tending to make it smaller the larger the VEV of the vortex
condensate is.  The resulting Casimir force can then be made sufficiently
small in the presence of vortex matter, though it will never be
exactly zero or become repulsive, as  it can be clear from the
expression for the Casimir force and from Eq. (\ref{m12}), where of
course $m_{1(2)}> 0$.


\section{Conclusions}
\label{sec5}

We have studied in this work how a nontrivial vacuum state,  with
condensed vortex excitations, affects the Casimir force  between two
conducting lines in a plane.  By starting from a CSH model with field
equations having vortex solutions, and using its dualized form, which
results to be a MCSH model, vortex degrees of freedom are made
explicit.  In the vortex condensation regime of the dual model, it can
be expressed simply as a MPCS theory, which in turn can be mapped in a
two noninteracting massive scalar field model.  Using the known
expression for the Casimir force for a massive scalar field, the
corresponding Casimir forces for the case of vortex matter between the
two lines have been computed.

We have shown that the Casimir force in the presence of vortex matter
is smaller than in the absence of vortices. This result may have
implications for Casimir effect experiments using
e.g. superconductors, like in the next generation of
experiments~\cite{next}, in the case that type-II superconductors
could eventually be used.  The results we have obtained are indicative
that the presence of vortices in the superconducting materials   can
make the Casimir effect much smaller, making its detection through
measurements more difficult.  Earlier experiments on the Casimir
effect performed by using superconducting materials, e.g. in
\cite{Bimonte1},  investigated the variation of the Casimir
energy in the transition from the normal to the superconducting state.
Though this variation can be very small, it can have a magnitude
comparable to the condensation energy of a semiconducting film. It has
been shown in \cite{Bimonte1} that this can cause a
measurable  increase in the value of the critical magnetic field
required for the  transition. However, these experiments were
performed by using type-I superconductors, where a vortex state is
absent.  It is feasible to expect, based on the results we have
obtained here, that in the case of type-II superconductors, there
should also be observed another  variation of the Casimir energy in
the transition from the superconducting  state to the Shubnikov phase,
where vortices are formed. 

Another important issue that must be cited is the possibility of using
our results to find the Casimir force, for the MPCS theory, in the
case of moving boundaries ({\it i.e.}, the dynamical Casimir
effect). As mentioned in the introduction, it is expected that the
Casimir energy plays an important role in superconductors, especially
at the nanometer scale. Recently, the first experimental observation of
the dynamical Casimir effect in a superconductor circuit \cite{Nature}
has brought great attention to this matter.  Some of the
considerations that we have done here are also valid in the dynamical
case.  Of course, where we set the boundaries (e.g. $x=0$ and $x = a$)
is of decisive importance for determining the expression for the
Casimir force.  However, the mapping between the initial MPCS theory
and the model of the scalar fields makes  use only of the values of
the derivatives of the functions  $\phi$, $\varphi$ and $A_i$ at the
boundaries. But the value of $x$ itself at those boundaries  is never
actually needed there at any step.  In other words, the mapping used
here is expected  also to be valid in the case of moving boundaries,
as long as the BCs remain valid (e.g., a perfect conductor parallel to
the $y$ axis, in a movement in the $x$ direction). Hence we conclude
that we can use the same arguments used here to study the dynamical
Casimir effect for the MPCS model. However, to find the Casimir force
in that case, we must know the force for a massive scalar field
between moving boundaries, which is an issue that we intend to treat
in a future work.


\appendix

\section{The Energy-Momentum tensor and the Casimir force}

The Casimir force for the dual theory is expressed, as usual, in terms
of the  VEV of the $T^{11}$ component of the symmetrized
energy-momentum tensor~\cite{Milton_1990,DanEdneyFelSil}:
$\mbox{force/length}  = \left\langle 0| T^{11} |0\right\rangle$. Thus,
we can first write

\begin{equation}
T^{\mu \nu } = F^{\mu}F^{\nu}+m^{2}A^{\mu }A^{\nu } -
\dfrac{1}{2}g^{\mu\nu}\left( F_{\lambda}F^{\lambda }+m^{2}A_{\alpha
}A^{\alpha }\right),
\end{equation}
where, for the sake of simplicity, we made use of the definition of
the dual tensor $F^{\mu}$,
\begin{equation}
\label{dual}
F^{\mu} \equiv \frac{1}{2}\epsilon^{\mu \nu \rho} F_{\nu \rho}
\end{equation}
Usually, the components $F^{\mu}$ are associated to the components of
the ``electric'' and ``magnetic" fields ($F^1 = -E_y$, $F^2 = E_x$ and
the scalar $B = F^0$). In this work, $E_x$, $E_y$ and $B$ may or may
not (in the case of the dual gauge field) represent a physical massive
electromagnetic field (we are just borrowing an usual nomenclature).
Hence, $\left\langle 0| T^{11} |0\right\rangle$ at the boundaries can
be written in terms of VEVs of products like $A^{\mu}A^{\nu}$ and
derivatives of them, taken at $x = 0$ or $x = a$ (the explicit values
of $x$ at the boundaries will not be necessary for our
purposes). {}Following \cite{Milton_1990}, we write those VEVs, at $x
= 0$, as 

\begin{equation}
\label{AmuAnuPropagators}
\left\langle 0| A^{\mu}(x)A^{\nu}(x) |0\right\rangle |_{x_1 = 0} =
\lim_{x_1 \to x'_1 = 0}\left\langle 0| A^{\mu}(x)A^{\nu}(x')
|0\right\rangle,
\end{equation} 
where $x$ and $x'$ stand for points in the three-dimensional
space-time. But the VEVs in the right-hand side 
of Eq. (\ref{AmuAnuPropagators})
are the two-point functions of the model, which can be written in
terms of the functional derivatives of the normalized generating
functional $Z[J_{\alpha}]$, where $J_{\alpha}$ is a source:

\begin{equation}
\label{derivativesOfZdeJ}
\left\langle 0| A^{\mu}(x)A^{\nu}(x') |0\right\rangle = -
\left. \frac{\delta^2 Z[J]}{\delta J_{\mu}(x)\ \delta J_{\nu}(x')}
\right|_{J_{\alpha} = 0} =  -\frac{\delta^2}{\delta J_{\mu}(x)\ \delta
  J_{\nu}(x')}\left[ \frac{\int {\cal D}A_{\beta} \exp \left(  iS +
    i\int J_{\alpha}A^{\alpha} \right)}{ \int {\cal D}A_{\gamma} \exp
    ( iS )}   \right]_{J_{\alpha} = 0}.
\end{equation} 
The derivatives appearing in Eq. (\ref{derivativesOfZdeJ}) are
independent of the norm of the field. This also shows that the Casimir
force should also be independent of the normalization of the
fields. {}For instance, the arbitrary mass parameter $\sigma$
appearing in Eq.~(\ref{S_ini})  can be put as a global multiplicative
constant $\sigma^2$  in all terms in Eq. (\ref{LMPCS}) and be
reabsorbed in a redefinition of the norm of $A_{\mu}$.


\begin{acknowledgments}
C.R.M.S. was supported by Coordena\c{c}\~ao de Aperfei\c{c}oamento de
Pessoal de N\'{i}vel Superior (CAPES-Brazil).  R.O.R. is partially
supported by Conselho Nacional de Desenvolvimento Cient\'{\i}fico e
Tecnol\'{o}gico (CNPq-Brazil) and by {}Funda{\c {c}}{\~{a}}o de Amparo
{\`{a}} Pesquisa do Estado do Rio de Janeiro (FAPERJ).
J.F.M.N. thanks J.A. Helay\"{e}l Neto and D. T. Alves for very
fruitful comments.
\end{acknowledgments}


\end{document}